\documentclass[submission, Phys]{SciPost}
\usepackage{bm}
\usepackage{bbm}
\usepackage[makeroom]{cancel}
\usepackage{nicefrac}
\usepackage{amsfonts}
\usepackage{pifont}

\usepackage{soul}  
\usepackage{hyperref}
\hypersetup{
    colorlinks=true,       
    linkcolor=blue,          
    citecolor=blue,       
    filecolor=magenta,      
    urlcolor=blue          
}

\newcounter{comm} 

\usepackage{adjustbox,expl3,etoolbox}
\letcs\replicate{prg_replicate:nn}
\newcommand*\longsum[1][1]{%
  \mathop{\textnormal{%
    \clipbox{0pt 0pt {.5\width} 0pt}{$\displaystyle\sum$}%
    \replicate{#1}{\clipbox{{.5\width} 0pt {.4\width} 0pt}{$\displaystyle\sum$}}%
    \clipbox{{.6\width} 0pt 0pt 0pt}{$\displaystyle\sum$}}}%
}

\newcommand{\eq}[1]{Eq.~(\ref{eq:#1})}
\newcommand{\eqs}[2]{Eqs.~(\ref{eq:#1}) and~(\ref{eq:#2})}

\newcommand{\eqr}[2]{Eqs.~(\ref{eq:#1}-\ref{eq:#2})}

\newcommand{\sref}[1]{Sec.~\ref{sec:#1}}

\newcommand{\aref}[1]{Appendix~\ref{sec:#1}}

\newcommand{\eql}[1]{\label{eq:#1}}

\newcommand{\secl}[1]{\label{sec:#1}}

\newcommand{\tick}{\ding{51}}
\newcommand{\cross}{\ding{55}}

\newcommand{\remove}[1]{}

\def\nn{\nonumber\\}

\def\beq{\begin{equation}}
\def\eeq{\end{equation}}
\def\bea{\begin{eqnarray}}
\def\eea{\end{eqnarray}}

\def\ip#1#2{\langle#1\vert#2\rangle}

\def\a{\alpha}
\def\b{\beta}
\def\c{\gamma}
\def\d{\delta}

\def\ab{{\a\b}}
\def\ba{{\b\a}}
\def\abc{{\a\b\c}}
\def\acb{{\a\c\b}}
\def\cba{{\c\b\a}}
\def\bca{{\b\c\a}}
\def\bac{{\b\a\c}}
\def\cab{{\c\a\b}}

\def\Aoit{a_0^+} 
\def\Aoti{a_0^-} 
\def\Aiot{a_1^-} 
\def\Atoi{a_1^+} 
\def\Aito{a_2^+} 
\def\Atio{a_2^-} 
\def\wAoit{\tilde{a}_0^+} 
\def\wAoti{\tilde{a}_0^-} 
\def\wAiot{\tilde{a}_1^-} 
\def\wAtoi{\tilde{a}_1^+} 
\def\wAito{\tilde{a}_2^+} 
\def\wAtio{\tilde{a}_2^-} 

\def\T{\mathcal{T}}

\def\kk{{\bm k}}
\def\MM{{\bm M}}
\def\jj{{\bm j}}

\def\OO{\mathbf{\Omega}}

\def\EE{{\bm E}}

\def\HH{\mathcal{H}} 
\def\OO{\mathcal{O}} 

\usepackage{multirow}
\usepackage{rotating}

\begin{document}

\begin{center}{\Large \textbf{On the separation of Hall and Ohmic
    nonlinear responses}}
\end{center}

\begin{center}
Stepan S. Tsirkin\textsuperscript{1*} and
Ivo Souza\textsuperscript{2,3}
\end{center}

\begin{center}
{\bf 1} Physik-Institut, Universit\"at
  Z\"urich, Winterthurerstrasse 190, CH-8057 Z\"urich, Switzerland
\\
{\bf 2} Centro de F{\'i}sica de Materiales,
  Universidad del Pa{\'i}s Vasco, 20018 San Sebasti{\'a}n, Spain
\\
{\bf 3} Ikerbasque Foundation, 48013 Bilbao, Spain
\\
\vspace{0.25cm}
* stepan.tsirkin@uzh.ch
\end{center}

\begin{center}
\today
\end{center}


\section*{Abstract} {\bf The symmetric and antisymmetric parts of the
  linear conductivity describe the dissipative (Ohmic) and
  nondissipative (Hall) parts of the current.  The Hall current is
  always transverse to the applied electric field regardless of its
  orientation; the Ohmic current is purely longitudinal in cubic
  crystals, but in lower-symmetry crystals it has a transverse
  component whenever the field is not aligned with a principal axis.
  In this work, we extend that analysis beyond the linear regime. We
  consider all possible ways of partitioning the current at any order
  in the electric field without taking symmetry into account, and find
  that the Hall vs Ohmic decomposition is the only one that satisfies
  certain basic requirements. A general prescription is given for
  achieving that decomposition, and the case of the quadratic
  conductivity is analyzed in detail. By performing a symmetry
  analysis we find that in five of the 122 magnetic point groups the
  quadratic dc conductivity is purely Ohmic and even under time
  reversal, a type of response that is entirely disorder mediated.}

\vspace{10pt}
\noindent\rule{\textwidth}{1pt}
\tableofcontents\thispagestyle{fancy}
\noindent\rule{\textwidth}{1pt}
\vspace{10pt}

\section{Introduction}
\label{sec:intro}
A static electric field applied to a conducting crystal generates a
current density that may be written to linear order as
\beq
j^{(1)}_\a = \sigma_\ab E_\b\,,
\eql{j1st}
\eeq
where a summation over Cartesian index $\b$ is implied, and
$\sigma_\ab$ is understood to be a function of the externally applied
magnetic field ${\bm H}$.  In general $\jj^{(1)}$ is not parallel to
$\EE$, but under certain conditions it may contain a part that is
always perpendicular to $\EE$, irrespective of how the field is
oriented relative to the crystal axes. This {\it Hall current} is
described by the antisymmetric part of the linear conductivity tensor,
\beq
j_{\HH,\a}^{(1)} = \sigma^\HH_\ab E_\b, \quad\quad
\sigma_\ab^\HH =
\frac{1}{2}\bigl( \sigma_\ab-\sigma_\ba \bigr)\,,
\eql{j1st-H}
\eeq
and the remainder $\jj^{(1)}_\OO=\jj^{(1)}-\jj^{(1)}_\HH$, given by
the symmetric part of the conductivity, is the {\it Ohmic current}
that gives rise to energy dissipation via Joule heating,
\beq
j_{\OO,\a}^{(1)} = \sigma^\OO_\ab E_\b, \quad\quad
\sigma_\ab^\OO =
\frac{1}{2}\bigl( \sigma_\ab+\sigma_\ba \bigr)\,.
\eql{j1st-O}
\eeq

Building on seminal works from the
1970s~\cite{baranova-oc77,ivchenko-jetp78,vorobev-jetp79}, there is at
present renewed interest in {\it nonlinear} effects in solids arising
from broken symmetries~\cite{tokura-nc18}. The nonlinear transport
effects that are being actively investigated include unidirectional
magnetoresistance (both induced by a magnetic
field~\cite{rikken-prl01,ideue-natphys17,Rikken-Te} and
spontaneous~\cite{avci-natphys15,olejnik-prb15,zelezny-prb21}), and
various nonlinear Hall
effects~\cite{deyo2009semiclassical,moore-prl10,gao-prl14,sodemann-prl15,nandy-prb19,ma-nature19,kang-natmater19,li-prb21,ortix2021nonlinear,du2021perspective,zhang2021higherorder,wang-prl21,liu-prl21,lai-natnano21}.

Despite the surge of interest in nonlinear currents, a clear
discussion of how to extend the Hall vs Ohmic decomposition to the
nonlinear regime is lacking, and confusing or even incorrect
statements can be found in the recent literature. With the present
work we aim to clarify the phenomenology of the nonlinear Hall vs
Ohmic decomposition, and to place it in the broader context of how to
partition the nonlinear current into physically well-defined parts.
Although we will focus on the conductivity tensor, our analysis
applies equally well to the resistivity. For simplicity, we will
assume throughout that the applied electric field is static (dc
limit).

To motivate the problem, consider the second-order response
\beq
j^{(2)}_\a =\sigma_{\a\b\c}E_\b E_\c\,,
\eql{j2nd}
\eeq
which requires broken inversion symmetry.  Contrary to the linear
conductivity, the quadratic conductivity is not uniquely defined since
adding to it a correction of the form
\begin{equation}
\Delta\sigma_\abc=-\Delta\sigma_\acb
\eql{gauge}
\end{equation}
does not change the physically observable current.  We will refer to
this freedom in defining nonlinear conductivities as a ``gauge
freedom,'' and to the unique choice that satisfies
$\sigma_\abc=\sigma_\acb$ as the ``symmetric gauge.'' Thus, the
symmetric gauge is the one where the conductivity tensor has intrinsic
permutation symmetry~\cite{boyd-book03}.

By analogy with \eq{j1st-H}, one might attempt to define
$\sigma^\HH_\abc$ as the part of $\sigma_\abc$ that is antisymmetric
in either the first and second indices,
\beq
\sigma^{1,2}_\abc=\frac{1}{2}\left( \sigma_\abc-\sigma_\bac\right)\,,
\eql{sigma-1-2}
\eeq
or in the first and third,
\beq
\sigma^{1,3}_\abc=\frac{1}{2}\left( \sigma_\abc-\sigma_\cba\right)\,.
\eql{sigma-1-3}
\eeq
(Note that we need to choose between these two options, since imposing
both conditions would render $\sigma^\HH_\abc$ totally antisymmetric,
resulting in zero current.)  Both choices yield Hall-like transverse
currents.  However, not only do they give different currents, but
those currents depend on the initial gauge choice for $\sigma_\abc$.
Both problems can be fixed by switching to the symmetric gauge,
$\overline{\sigma}_\abc=\frac{1}{2}\left(
\sigma_\abc+\sigma_\acb\right)$, before applying the
antisymmetrization \eqref{eq:sigma-1-2} or \eqref{eq:sigma-1-3}.
Since the resulting Hall-like conductivities
\beq
\overline{\sigma}^{1,2}_\abc=\frac{1}{4}
\left(\sigma_\abc+\sigma_\acb-\sigma_\bac-\sigma_\bca\right)
\eql{sigma-1-2-bar}
\eeq
and
\beq
\overline{\sigma}^{1,3}_\abc=\frac{1}{4}
\left(\sigma_\abc+\sigma_\acb-\sigma_\cba-\sigma_\cab\right)
\eql{sigma-1-3-bar}
\eeq
satisfy $\overline{\sigma}^{1,3}_\abc=\overline{\sigma}^{1,2}_\acb$,
they clearly yield the same current (they are related by the gauge
transformation
$\Delta\sigma_{\a\b\c}=\overline{\sigma}^{1,2}_{\a\c\b}-\overline{\sigma}^{1,2}_{\a\b\c}$). This
modified prescription~\cite{nandy-prb19,ortix2021nonlinear} is
nevertheless still not quite correct.

As a concrete example, we take the expression for the quadratic
conductivity obtained by solving the Boltzmann equation at
${{\bm H}}={\bf 0}$ in the constant relaxation-time approximation.
Denoting the relaxation time as $\tau$, there are
contributions of order $\tau^0$, $\tau^1$, and $\tau^2$, with those of
even (odd) order in $\tau$ being odd (even) under time reversal
$\T$~\cite{gao-fp19} . Neglecting disorder-mediated contributions
(skew-scattering and side jump) one
finds~\cite{deyo2009semiclassical,gao-prl14,sodemann-prl15,zhang2021higherorder,zelezny-prb21}
\beq
\sigma_{\abc}=\frac{e^3}{\hbar}\int_{\kk n}\,f_0(\epsilon_n)
\left[
\left(\partial_\a G_n^{\b\c}-\partial_\b G_n^{\a\c}\right)+
(\tau/\hbar)\partial_\c\Omega_n^{\a\b}-
(\tau/\hbar)^2\partial^3_{\a\b\c}\epsilon_n
\right]\,,
\eql{sigma-abc-tot}
\eeq
where $\int_{\kk n}\equiv d^d k/(2\pi)^d\sum_n$ in $d$ dimensions and
we have dropped $\kk$ from the integrand, $e>0$ is the elementary
charge, $\epsilon_n$ is the band energy, $f_0$ is the Fermi-Dirac
distribution function, and $\partial_\c\equiv\partial/\partial k_\c$.
$\Omega_n^{\a\b}$ is the Berry curvature, and $G_n^{\a\b}$ is
sometimes called the Berry curvature polarizability; these two
quantities can be expressed in terms of the Berry connection matrix
$A_{mn}^\a=i\ip{u_m}{\partial_\a u_n}$ as follows,
\beq
\Omega_n^{\a\b}=\partial_\a A_{nn}^\b-\partial_\b A_{nn}^\a=
-2\text{Im}\ip{\partial_\a u_n}{\partial_\b u_n}\,\,,
\eeq
\beq
G_n^{\a\b}=-2\text{Re}\,\sum_m^{\epsilon_m\not=\epsilon_n}\,
\frac{A^\a_{nm}A^\b_{mn}}{\epsilon_n-\epsilon_m}\,.
\eeq
The ${\cal O}(\tau^0)$ and ${\cal O}(\tau^1)$ terms in in
\eq{sigma-abc-tot} describe respectively $\T$-odd and $\T$-even
quadratic anomalous Hall responses whose net current we denote by
${\bm j}^{(2)}_\HH$, and the ${\cal O}(\tau^2)$ term is a $\T$-odd
Drude-like quadratic conductivity that has been identified as a
mechanism for spontaneous unidirectional
magnetoresistance~\cite{zelezny-prb21}.  Applying to
\eq{sigma-abc-tot} each of the prescriptions in
\eqr{sigma-1-2}{sigma-1-3-bar}, we obtain
\beq
\left({\bm j}^{1,2},{\bm j}^{1,3},
\overline{\bm j}^{1,2}=\overline{\bm j}^{1,3}\right)=
\left(1,\nicefrac{1}{2},\nicefrac{3}{4}\right){\bm j}^{(2)}_\HH
\eql{hall-currents-proposed}
\eeq
for the quadratic Hall currents.  Prescription~\eqref{eq:sigma-1-2}
gives the full Hall current ${\bm j}^{(2)}_\HH$, but that is
accidental: if we make the gauge transformation
$\sigma_\abc\rightarrow\sigma_\acb$ in \eq{sigma-abc-tot}, the Hall
currents obtained from prescriptions~\eqref{eq:sigma-1-2}
and~\eqref{eq:sigma-1-3} get swapped:
$\left({\bm j}^{1,2},{\bm j}^{1,3}\right)\rightarrow
\left(\nicefrac{1}{2},1\right){\bm j}^{(2)}_\HH$. We mentioned earlier
that the prescriptions in \eqs{sigma-1-2-bar}{sigma-1-3-bar} are not
quite correct, and indeed they only recover three quarters of the full
Hall current; we will see in \sref{second-order} that multiplying the
right-hand sides of those equations by factors of $4/3$ does lead to
generally valid expressions for the quadratic Hall conductivity.

The strategies in \eqr{sigma-1-2}{sigma-1-3-bar}, which constitute
attempts to generalize to third-rank tensors the definition in
\eq{j1st-H} of an antisymmetric tensor of rank two, fail to yield a
proper decomposition of the quadratic current.  On the other hand,
higher-order generalizations of the symmetrization procedure in
\eq{j1st-O} are straightforward, since one can symmetrize over all
indices. In the case of the quadratic conductivity one finds
\beq
\sigma^\OO_{\a\b\c}=\frac{1}{6}
\left(
\sigma_{\a\b\c}+\sigma_{\a\c\b}+\sigma_{\b\a\c}+
\sigma_{\b\c\a}+\sigma_{\c\a\b}+\sigma_{\c\b\a}
\right)\,,
\eql{sigma-symmetric}
\eeq
and it can be readily checked that the power dissipation is fully
accounted for by $\sigma^\OO_{\a\b\c}$,
\beq
{\bm j}^{(2)}\cdot\EE=\sigma_\abc E_\a E_\b E_\c=
\sigma^\OO_\abc E_\a E_\b E_\c\,,
\eql{dissipation-2nd}
\eeq
which justifies calling it the quadratic Ohmic
conductivity. Accordingly,
\beq
\sigma^\HH_\abc=\sigma_\abc-\sigma^\OO_\abc
\eql{sigma-hall}
\eeq
describes the dissipationless (Hall) part of the quadratic current
response.

Surprisingly we could not find, in the growing literature on nonlinear
currents in solids, any explicit mention of the simple prescription in
\eqs{sigma-symmetric}{sigma-hall} for separating the nonlinear Hall
and Ohmic conductivities. Let us apply it to the expression in
\eq{sigma-abc-tot} for the quadratic conductivity.  Since the
${\cal O}(\tau^0)$ and ${\cal O}(\tau^1)$ terms therein are
antisymmetric in two indices, they drops out from
\eq{sigma-symmetric}; and since the {${\cal O}(\tau^2)$} is already
totally symmetric, it becomes the full $\sigma^\OO_{\a\b\c}$.  Hence,
the former terms are Hall-like and the latter is Ohmic.

It should be noted that we have not yet proven that
\eqs{sigma-symmetric}{sigma-hall} give the \textit{only} valid
decomposition of the quadratic current into Ohmic and Hall parts.  For
example, one could define another partition
\beq
\tilde\sigma^\HH_\abc = (1-x) \sigma^\HH_\abc\;,  \quad\quad
\tilde\sigma^\OO_\abc = \sigma^\OO_\abc +x \sigma^\HH_\abc   \quad\quad
(x\in\mathbb{R})
\eeq
that is not related to that of \eqs{sigma-symmetric}{sigma-hall} by
any gauge transformation~{\eqref{eq:gauge}}, and again
$\tilde\sigma^\HH_\abc$ would describe a dissipationless current, with
all the Joule heating coming from $\tilde\sigma^\OO_\abc$.

In this work, we consider the problem of defining nonlinear Hall
and Ohmic conductivities from a more general perspective. Our starting
point is the following question:\\

\textit{What are all the possible ways of partitioning the nonlinear
  current into physically meaningful parts, without taking
  into account {neither} the symmetries of the system nor specific microscopic mechanisms?}\\

\noindent
(We will refer to such partitions as ``generic.'') To address this
question, we start by formulating in \sref{general} the necessary
criteria for a proper generic partition of the current at arbitrary
order in $\EE$.  In \sref{second-order} we find that there is a
\textit{unique} nontrivial decomposition of the current at second
order that fulfils those criteria, which corresponds precisely to the
Hall vs Ohmic decomposition.  (Our criteria do not single out any
particular gauges for the partial nonlinear conductivities; instead,
they take the form of necessary and sufficient conditions satisfied by
the partial conductivities in arbitrary gauges.)  The Hall vs Ohmic
decomposition is generalized to arbitrary order in
\sref{higher-order}. In \sref{symmetry} we return to the quadratic
conductivity to carry out a systematic symmetry analysis of its Hall
and Ohmic parts, and in \sref{conclusions} we draw conclusions. In
\aref{general-proof} we prove that the Hall vs Ohmic partition of the
current is the only generic partition possible at every order in
$\EE$, and in \aref{repackaging} we repackage the disorder-free
quadratic conductivity~\eqref{eq:sigma-abc-tot} in the manner
described in \sref{symmetry}.


\section{Criteria for a
  generic  partition of the current
}
\secl{general}
Our strategy for partitioning the nonlinear current will be as
follows.  We start from a conductivity tensor
$\sigma_{\a_0\a_1\ldots\a_{n}}$ describing the full $n$-th order
response,
\beq
j^{(n)}_{\a_0}=\sigma_{\a_0\a_1\ldots\a_n}{E_{\a_1}\ldots E_{\a_n}}\,,
\eql{jnth}
\eeq
and search for an operator $\hat P$ that selects part of this
current. We want the operator $\hat P$ to act order by order in the
electric field; this means that its action on the full conductivity
tensor should result in a linear combination of versions of that same
tensor with different sets of indices,
\beq
\left(\hat{P}\sigma\right)_{\a_0\a_1\ldots\a_n}
=\sum_{p}\,{a_p} \sigma_{\a_{p(0)}\a_{p(1)}\ldots\a_{p(n)}}\,.
\eql{Psigma}
\eeq
Here the summation is over all possible mappings
\beq
\{0,1,\ldots,n\}\,\overset{p}{\longrightarrow}\,\{ p(0),p(1),\ldots,p(n)\}
\eql{mapping}
\eeq
where $p(n)\in\{0,1,\ldots,n\}$, and $a_p$
are
coefficients to be determined.
The part of the current  selected by $\hat{P}$
can be written symbolically as
\beq
\left(\hat{P}\jj^{(n)}\right)_{\alpha_0} =
\left(\hat{P}\sigma\right)_{\a_0\a_1\ldots\a_n}{E_{\a_1}\ldots E_{\a_n}}\,.
\eql{Pj}
\eeq

We shall require three properties of $\hat{P}$.  The first is that
it acts on the current as a projector, so that

\beq
\hat{P}\jj^{(n)}=\hat{P}^2\jj^{(n)}\,;
\eql{idemp}
\eeq
the second is that the projected current is invariant under gauge
transformations of the full $n$-th order conductivity tensor, that is,
\beq
\Delta\Big(\hat{P}\jj^{(n)}\Big)=0
\eql{gauge-inv}
\eeq
whenever $\Delta\jj^{(n)}=0$, which in turn holds if and only if
$\Delta\sigma_{\a_0\a_1\ldots\a_n}$ vanishes under symmetrization over
the last $n$ indices.

Finally, we require that the projected current transforms as a vector
under rotations of the coordinate system, so that
$\hat{P}\jj^{(n)}\cdot\EE$ remains invariant under such
transformations. This is justified by the intention to arrive at a
generic prescription that is not bound to any particular crystal
symmetry, and not even to a specific number of spatial dimensions.
This third constraint will be satisfied if the summation in
\eq{Psigma} is restricted to \textit{permutation} mappings $p$, for
which $p(i)\not= p(j)$ whenever $i\not= j$.  Conversely, if mappings
with $p(i)=p(j)$ for some $i\not=j$ are included, scalar products will
not be conserved under rotations.\footnote{Take for example
  $\hat{P}_\HH\sigma_{\a\b}=(\sigma_{\a\a}+\sigma_{\b\b})/2$.  For an
  electric field lying on the $xy$ plane this gives
  $\hat{P}_\HH\jj^{(1)}\cdot\EE =
  E_x^2\sigma_{xx}+E_y^2\sigma_{yy}+E_x E_y
  (\sigma_{xx}+\sigma_{yy})$, and the result should be the same in a
  different coordinate system.  However, in a coordinate system that
  differs by a two-fold rotation about the $y$ axis we obtain
  $\hat{P}_\HH\jj^{(1)}\cdot\EE = E_x^2\sigma_{xx}+E_y^2\sigma_{yy} -
  E_x E_y (\sigma_{xx}+\sigma_{yy})$, which is a different result.}
Thus, from here on we shall restrict our attention to permutation
mappings, and investigate which operators $\hat{P}$ can satisfy the
two conditions expressed by \eqs{idemp}{gauge-inv}.

Before proceeding, we note that if we find some operator $\hat{P}$
that satisfies the conditions listed above, those conditions will also
be satisfied by $\hat{P}^\prime = \hat{1}-\hat{P}$.  Thus, any
nontrivial operator $\hat{P}$ defines a decomposition of the current
into two parts (by ``nontrivial'' we mean an operator such that
$\hat{P}\jj\neq{\bf 0}$ and $\hat{P}\jj\neq\jj$).  We will start by
applying the above criteria to the second-order response, and then we
will generalize to higher orders.


\section{Second-order response}
\secl{second-order}
Consider an operator $\hat{P}$ acting on the quadratic conductivity
according to \eq{Psigma},
\beq
\hat{P}\sigma_\abc = \Aoit\sigma_\abc + \Aoti\sigma_\acb
                   + {\Atoi\sigma_\cab + \Aiot\sigma_\bac
                   + \Aito\sigma_\bca} + \Atio\sigma_\cba\,,
\eql{P-2nd}
\eeq
and on the quadratic current according to \eq{Pj},
\beq
\hat{P}j^{(2)}_\a=\left(\hat{P}\sigma_\abc\right)E_\b E_\c
=\left(A_0\sigma_{\a\b\c}+A_1\sigma_{\b\a\c}+
A _2\sigma_{\b\c\a}\right)E_\b E_\c\,.
\eql{Pj-2nd}
\eeq
Here $A_i={a_i^+} + {a_i^-}$, and the notation for the coefficients
$a_i^\pm$ is as follows: the subscript denotes the position of
$\alpha$ in the permutation of the indices $\abc$, and the superscript
gives the parity of the permutation.

Our claim is that $\hat{P}$ yields a proper generic partition of the
current only if it satisfies \eqs{idemp}{gauge-inv}.  Let us start
with the gauge-invariance condition~\eqref{eq:gauge-inv}.  The
projected current~\eqref{eq:Pj-2nd} remains unchanged under the gauge
transformation~\eqref{eq:gauge} if and only if
\beq
(A_1-A_2)E_\b E_\c\Delta\sigma_{\b\a\c}= 0\,;
\eql{gauge-inv-equation-2nd}
\eeq
since this condition must be satisfied for arbitrary $\EE$, and since we did
not set any rules for permutations involving the first index of
$\Delta\sigma$, it follows that $A_1=A_2$.  To impose the idempotency
condition~\eqref{eq:idemp}, we first apply \eq{P-2nd} recursively to
find
\beq
\hat{P}^2\sigma_\abc
                   = \wAoit\sigma_\abc + \wAoti\sigma_\acb
                   + \wAtoi\sigma_\cab + \wAiot\sigma_\bac
                   + \wAito\sigma_\bca + \wAtio\sigma_\cba\,,
\eql{P2-2nd}
\eeq
where
\begin{subequations}
\begin{align}
\wAoit&=\Aoit\Aoit+\Aoti\Aoti+\Aiot\Aiot+2\Aito\Atoi 
+\Atio\Atio\,,\\
\wAoti&=2\Aoit\Aoti+\Atoi\Aiot+\Atoi\Atio+\Aiot\Aito+\Aito\Atio\,,\\
{\wAtoi}&=2\Aoit\Atoi+\Aoti\Aiot+\Aoti\Atio+\Aiot\Atio+\Aito\Aito\,,\\
{\wAiot}&=2\Aoit\Aiot+\Aoti\Atoi+\Aoti\Aito+\Atoi\Atio+\Aito\Atio\,,\\
{\wAito}&=2\Aoit{\Aito+\Aoti\Aiot+\Aoti\Atio+\Atoi\Atoi+\Aiot\Atio}\,,\\
\wAtio&=2\Aoit\Atio+\Aoti\Atoi+\Aoti\Aito+\Atoi\Aiot+\Aiot\Aito\,.\
\end{align}
\eql{Atilde}
\end{subequations}
By analogy with \eq{Pj-2nd} we have
\beq
\hat{P}^2j^{(2)}_\a=
\left(\tilde{A}_0\sigma_{\a\b\c}+\tilde{A}_1\sigma_{\b\a\c}+
\tilde{A} _2\sigma_{\b\c\a}\right)E_\b E_\c
\eql{P2j-2nd}
\eeq
for the twice-projected current, where the coefficients
$\tilde{A}_i=\tilde{a}_i^+ + \tilde{a}_i^-$ are given by
\begin{subequations}
\begin{align}
\tilde{A}_0&=A_0^2\quad\;+ { (\Aiot+\Aito)A_1+(\Atoi+\Atio)A_2}\,,\\
\tilde{A}_1&=A_0 A_1+(\Aoit+\Atio)A_1+(\Aoti+\Aito)A_2\,,\\
\tilde{A}_2&=A_0 A_2+(\Aoti+\Atoi)A_1+(\Aoit+\Aiot)A_2\,.
\end{align}
\eql{A0A1A2tilde}
\end{subequations}
Equating~\eqref{eq:Pj-2nd} and~\eqref{eq:P2j-2nd}, the idempotency
condition becomes $A_i=\tilde{A}_i$ for $i=0,1,2$.  Substituting
\eq{A0A1A2tilde} for $\tilde{A}_i$ and then invoking the gauge invariance
condition $A_1=A_2$, we are left with two conditions only,
\beq
A_0=A_0^2+2A_1^2\,,\quad\quad
A_1=(2A_0+A_1)A_1\,.
\eql{idemp-equations-2nd}
\eeq
These equations have four solutions. There are two solutions with
$A_1=0$,
\beq
\begin{cases}
\hat{P}_0:  (A_0,A_1=A_2)=(0,0)\\
\hat{P}_1:  (A_0,A_1=A_2)=(1,0)\\
\end{cases}{\quad\Rightarrow\quad
\jj^{(2)}={\bf 0}+\jj^{(2)}}\,,
\eql{trivial}
\eeq
which as indicated give the trivial ``all or nothing'' partition of
the current.  Then there are two solutions with $A_1\not=0$,
\beq
\begin{cases}
\hat{P}_\HH:  (A_0,A_1=A_2)=(\frac{2}{3},-\frac{1}{3})\\
\hat{P}_\OO:  (A_0,A_1=A_2)=(\frac{1}{3},\frac{1}{3})\\
\end{cases}{\,\Rightarrow\;
\jj^{(2)}=\jj^{(2)}_\HH+\jj^{(2)}_\OO}\,,
\eql{nontrivial}
\eeq
which give the desired Hall vs Ohmic partition.  To show that this is
the case, we turn to the condition that defines a Hall-like projected
current,
\beq
\hat{P}\jj^{(2)}\cdot\EE=0\,,\quad\forall\EE\,. 
\eql{Hall-like-2nd}
\eeq
Using \eq{Pj-2nd} that condition becomes
$A_0+A_1+A_2=0$, which is satisfied by $\hat{P}_\HH$ but not by
$\hat{P}_\OO$.
This conclude the proof that \eqs{idemp}{gauge-inv} lead to a
partition of the quadratic current into Hall and Ohmic parts.
Remarkably, we found that this is in fact the only gauge-invariant and
idempotent generic partition possible, apart from the trivial one in
\eq{trivial}.

Since we are still free to adjust the six coefficients $a_i^\pm$ in
\eq{P-2nd} as long as $A_i=a_i^+ + a_i^-$ maintain the values given in
\eq{nontrivial}, the Hall and Ohmic quadratic conductivities are
highly nonunique.  This nonuniqueness corresponds precisely to the
gauge freedom~\eqref{eq:gauge} in defining $\sigma^\HH_{\abc}$ and
$\sigma^\OO_{\abc}$, and it does not affect the physical currents
$\jj^{(2)}_\HH$ and $\jj^{(2)}_\OO$. One way to fulfill the ``Ohmic''
conditions in \eq{nontrivial} is by setting all six coefficients in
\eq{P-2nd} to $1/6$, which leads to the fully symmetric form for
$\sigma^\OO_{\a\b\c}$ in \eq{sigma-symmetric}.

Let us now revisit the prescriptions proposed in
\eqs{sigma-1-2-bar}{sigma-1-3-bar} for defining $\sigma^\HH_{\a\b\c}$,
which consist in first symmetrizing the full $\sigma_{\abc}$ in the
last two indices, and then antisymmetrizing the first index with
either the second or the third~\cite{nandy-prb19,ortix2021nonlinear}.
When applied to a concrete example in \sref{general}, those
prescriptions only recovered three quarters of the full Hall current
[see \eq{hall-currents-proposed}]. This suggests it may be possible to
fix them by multiplying each of \eqs{sigma-1-2-bar}{sigma-1-3-bar} by
a factor of $4/3$,
\beq
\overline{\sigma}^{\HH(1,2)}_\abc=\frac{4}{3}\overline\sigma^{1,2}_\abc=
\frac{1}{3}
\left(\sigma_\abc+\sigma_\acb-\sigma_\bac-\sigma_\bca\right)\,,
\eql{sigma-1-2-tilde}
\eeq
\beq
\overline{\sigma}^{\HH(1,3)}_\abc=\frac{4}{3}\overline\sigma^{1,3}_\abc=
\frac{1}{3}
\left(\sigma_\abc+\sigma_\acb-\sigma_\cba-\sigma_\cab\right)\,.
\eql{sigma-1-3-tilde}
\eeq
Comparing with \eq{P-2nd} we find
\beq
\Aoit=\Aoti=-\Aiot=-\Aito=\frac{1}{3}\,,\quad,\Atoi=\Atio=0
\eeq
in the case of \eq{sigma-1-2-tilde}, and
\beq
\Aoit=\Aoti=-\Atoi=-\Atio=\frac{1}{3}\,,\quad,\Aiot=\Aito=0
\eeq
in the case of \eq{sigma-1-3-tilde}.  Since both sets of coefficients
satisfy the Hall-like conditions in \eq{nontrivial},
\eqs{sigma-1-2-tilde}{sigma-1-3-tilde} are generally valid expressions
for the quadratic Hall conductivity.


\section{Higher-order responses}
\secl{higher-order}
At $n$-th order in the electric field, the Ohmic conductivity can be
chosen as the fully symmetrized conductivity tensor obtained by
setting $a_p=1/(n+1)!$ for all $p$ in \eq{Psigma},
\beq
\sigma^\OO_{\a_0\a_1\ldots\a_n}
\equiv\hat{P}_\OO\sigma_{\a_0\a_1\ldots\a_n}
=\frac{1}{(n+1)!}\sum_p\,
\sigma_{\a_{p(0)}\a_{p(1)}\ldots \a_{p(n)}}\,.
\eql{POn}
\eeq
This generalizes to arbitrary $n$ the symmetrization procedure of
\eqs{j1st-O}{sigma-symmetric} for $n=1$ and $n=2$, respectively.

Let us now show that with the above choice of Ohmic projector, the
Hall projector $\hat{P}_\HH=\hat{1}-\hat{P}_\OO$ satisfies
\eqs{idemp}{gauge-inv}.  We start again with the gauge invariance
condition. Since the full $n$-th order current is by definition
invariant under a gauge transformation
$\Delta\sigma_{\alpha_0\alpha_1\ldots\alpha_n}$, the Hall part is
invariant if and only if the Ohmic part is invariant. It is therefore
sufficient to show that
\beq
\Delta\left(\hat{P}_\OO j^{(n)}_{\a_0}\right)=
\left(
\hat{P}_\OO\Delta\sigma_{\a_0\a_1\ldots\a_n}
\right)
E_{\a_1}\ldots E_{\a_n}
\eeq
vanishes for arbitrary $\EE$. But since
$\Delta\sigma_{\alpha_0\alpha_1\ldots\alpha_n}$ must vanish under
symmetrization over the last $n$ indices to ensure that
$\Delta\jj^{(n)}={\bf 0}$ (see \sref{general}), it also vanishes under
full symmetrization by $\hat{P}_\OO$.  Next, it is clear that
$\hat{P}_\OO^2\sigma_{\a_0\a_1\ldots\a_n} =
\hat{P}_\OO\sigma_{\a_0\a_1\ldots\a_n}$
because symmetrization of tensor that is already fully symmetric does
not change it further. Therefore,
\beq
\hat{P}_\HH^2\jj^{(n)} = \left(1-2\hat{P}_\OO+\hat{P}_\OO^2\right)\jj^{(n)}
=\left(1-\hat{P}_\OO\right)\jj^{(n)} = \hat{P}_\HH\jj^{(n)}\,. 
\eeq
Finally, from the $n$-th order generalization of \eq{dissipation-2nd}
it follows that
$\jj^{(n)}_\HH=\jj^{(n)}-\jj^{(n)}_\OO$ is dissipationless.
Thus we have obtained a solution that satisfies \eqs{idemp}{gauge-inv}
at any order in $\EE$, and found that it corresponds to the Hall vs
Ohmic partition of the current.

To recapitulate, one can always define the Ohmic part of the $n$-th
order conductivity as the totally symmetric part, and the Hall part as
the remainder. For $n=1$, this procedure reduces to the standard
partition of the linear conductivity according to
\eqs{j1st-H}{j1st-O}.  We demonstrated in \sref{second-order} that for
$n=2$ the same procedure leads to the only well-defined (idempotent
and gauge-invariant) generic partition of the quadratic current, and
in \aref{general-proof} we generalize that proof to arbitrary~$n$.


\section{Symmetry analysis of the
  quadratic dc conductivity}
\secl{symmetry}
At linear order in $\EE$, the Hall vs Ohmic decomposition is
intimately related to time-reversal symmetry $\T$ by virtue of the
Onsager reciprocity relation
\begin{equation}
\sigma_\ab({\bm H},\MM)  =  \sigma_\ba(-{\bm H},-\MM)\,.
\eql{onsager}
\end{equation}
It follows from this relation that the Ohmic part of the linear
response is $\T$-even, while the Hall part is
$\T$-odd~\cite{SHTRIKMAN1965147,grimmer1993}.  In the nonlinear
regime, both Hall and Ohmic responses can have $\T$-even and $\T$-odd
components; this gives four contributions in total, of which only
three are present in \eq{sigma-abc-tot} for the disorder-free
$\sigma_\abc$. The reason why there is no $\T$-even Ohmic term in
\eq{sigma-abc-tot} is that in the semiclassical wavepacket formalism
there is no correction to the band energy at first order in the
electric field~\cite{xiao-rmp10}; the leading correction is of second
order, and it contributes to the $\T$-even cubic
conductivity~\cite{lai-natnano21}.

We will see shortly that $\sigma_\abc$ is purely Ohmic and $\T$-even
in five of the 122 magnetic point groups. Since for materials in those
point groups the disorder-free part of $\sigma_\abc$ vanishes
identically, their symmetry-allowed quadratic response must be
entirely disorder-mediated; this is consistent with the finding that a
skew-scattering contribution to $\sigma_\abc$ is present in all
non-centrosymmetric materials~\cite{isobe-sciadv20}. Contributions
from disorder to $\sigma^\HH_\abc$ have been studied
recently~\cite{konig-prb19,du-natcomms19,nandy-prb19}, but similar
contributions to $\sigma^\OO_\abc$ have received little attention so
far. In this regard, we note that the expressions for $\sigma_\abc$
obtained in
Refs.~\cite{konig-prb19,du-natcomms19,nandy-prb19,isobe-sciadv20}
contain not only Hall but also Ohmic parts, which can be separated out
using \eqs{sigma-symmetric}{sigma-hall}.

In preparation for performing a symmetry analysis of $\sigma_\abc$,
let us count the number of independent coefficients needed to describe
the quadratic Ohmic and Hall responses in two-dimensional (2D) and
three-dimensional (3D) space.  As $\sigma_\abc$ can be chosen to be
symmetric in the last two indices, it has 6 (18) independent
components in 2D (3D). $\sigma^\OO_\abc$ can be chosen to be fully
symmetric, and hence it has 4 (10) independent components in 2D (3D),
leaving $\sigma^\HH_\abc$ with $6-4=2$ ($18-10=8$) independent
components in 2D (3D).  Those Hall-like components can be repackaged
as an axial vector in 2D, and as a traceless rank-2 axial tensor in
3D. Choosing the latter as
\beq
\chi^{\HH}_{\c\d}=
\frac{1}{2}\varepsilon_{\a\b\c}\overline{\sigma}^{\HH(1,2)}_{\a\b\d}=
\frac{1}{2}\varepsilon_{\a\b\c}\overline{\sigma}^{\HH(1,3)}_{\a\d\b}
\eql{chi-H-a}
\eeq
and using either \eq{sigma-1-2-tilde} or \eq{sigma-1-3-tilde}, one
finds
\beq
\chi^{\HH}_{\c\d}=\frac{1}{3}\varepsilon_{\abc}
\left(\sigma_{\a\b\d}+\sigma_{\a\d\b}\right)\,.
\eql{chi-H}
\eeq
The tensor $\chi^{\HH}$ is traceless,\footnote{The fact that
  $\chi^{\HH}$ is traceless went unnoticed in Ref.~\cite{nandy-prb19},
  where it is stated that $\chi^{\HH}$ has nine independent components
  rather than eight.} and it remains invariant under gauge
transformations of the quadratic conductivity [\eq{gauge}].  This
gauge-invariant repackaging of the quadratic Hall conductivity tensor
is analogous to the repackaging
$\chi^\HH_\c=\varepsilon_{\c\a\b}\sigma_{\a\b}/2$ of the linear Hall
conductivity as an axial vector. As an example, in \aref{repackaging}
we evaluate $\chi^{\HH}$ for the disorder-free quadratic
conductivity~\eqref{eq:sigma-abc-tot}.

\begin{table}
\centering
\begin{tabular}{c|cc}
& Quadratic Ohmic & Quadratic Hall\\ 
\hline \\
$\T$-even & $[\text{V}^3]$ & $\text{eV}^2$ (traceless part)\\ \\
$\T$-odd & $\text{a}[\text{V}]^3$ & $\text{aeV}^2$ (traceless part)\\
\end{tabular}
\caption{Decomposition of the quadratic conductivity into Ohmic vs
  Hall parts and $\T$-even vs $\T$-odd parts. The Ohmic part is
  represented by a totally symmetric rank-3 polar tensor
  [\eq{sigma-symmetric}], and the Hall part by a traceless rank-2
  axial tensor [\eq{chi-H}]. Each entry in the table denotes the
  corresponding Jahn symbol~\cite{jahn-actacrys49}.}
\label{tab:4-parts}
\end{table}

According to the preceeding analysis, the quadratic conductivity can
be divided quite generally into an Ohmic part given by a totally
symmetric rank-3 polar tensor [\eq{sigma-symmetric}] and a Hall part
expressible as a traceless rank-2 axial tensor [\eq{chi-H}]. Each of
these can be further decomposed into $\T$-even and $\T$-odd parts,
resulting in a total of four contributions whose Jahn
symbols~\cite{jahn-actacrys49} are indicated in
Table~\ref{tab:4-parts}.

\begin{table}[h]
\begin{tabular}{p{8.1cm}cccc}
\hline\hline
& \multicolumn{2}{c}{Quadratic Hall} & \multicolumn{2}{c}{Quadratic Ohmic}\\
Magnetic point groups & $\T$-odd & $\T$-even &
$\T$-odd & $\T$-even\\
\hline
%
$-1$, $-11'$, 2/m, 2/m1$'$, 2$'$/m$'$, mmm, mmm1$'$, m$'$m$'$m, 4/m, 4/m1$'$,
4$'$/m, 4/mmm,
4/mmm1$'$, 4$'$/mm$'$m, 4/mm$'$m$'$, $-3$, $-31'$, $-3$m, $-3$m1$'$, $-3$m$'$,
6/m, 6/m1$'$, 6$'$/m$'$, 6/mmm, 6/mmm1$'$, 6$'$/m$'$mm$'$, 6/mm$'$m$'$,
m$-3$, m$-31'$, 432, $4321'$, m$-3$m, m$-3$m1$'$, m$-3$m$'$, m$'-3'$m$'$ &
\cross & \cross & \cross & \cross\\
\hline
4/m$'$m$'$m$'$, 6/m$'$m$'$m$'$ &
\tick & \cross & \cross & \cross\\
4221$'$, 6221$'$ &
\cross & \tick & \cross & \cross\\
422, 622 &
\tick & \tick & \cross & \cross\\
\hline
%
6$'$/m, 6$'$/mmm$'$, m$'-3'$, $4'32'$, m$'-3'$m &
\cross & \cross & \tick & \cross\\
$-61'$, $-6$m21$'$, 231$'$, $-43$m1$'$, $-4'3$m$'$ &
\cross & \cross & \cross & \tick\\
$-6$, $-6$m2, -6m$'2'$, 23, $-43$m &
\cross & \cross & \tick & \tick\\
\hline
$-6'$m$'$2 &
\tick &\cross & \cross & \tick\\
$6'22'$ &
\cross & \tick & \tick & \cross\\
$-1'$, 2$'$/m, 2/m$'$, m$'$mm, m$'$m$'$m$'$, 4/m$'$, 4$'$/m$'$, 4/m$'$mm,
4$'$/m$'$m$'$m, $-3'$, $-3'$m, $-3'$m$'$, 6/m$'$, 6/m$'$mm &
\tick & \cross & \tick & \cross\\
$11'$, $21'$, m$1'$, 2221$'$, mm21$'$, 41$'$, $-41'$, 4mm1$'$, $-42$m1$'$,
31$'$, $321'$, 3m1$'$, 61$'$, 6mm1$'$ &
\cross & \tick & \cross & \tick\\
$4'22'$, $42'2'$, $62'2'$ &
\tick & \tick & \tick & \cross \\ 
4m$'$m$'$, $-4'2$m$'$, 6m$'$m$'$ &
\tick & \tick &\cross & \tick\\
$-6'$, $-6'$m2$'$ &
\tick & \cross & \tick & \tick\\
6$'$, 6$'$mm$'$ &
\cross & \tick & \tick & \tick\\
1, 2, 2$'$, m, m$'$, 222, $2'2'2$, mm2, m$'$m2$'$, m$'$m$'$2, 4, 4$'$,$-4$,
$-4'$, 4mm, 4$'$m$'$m,
$-42$m, $-4'2'$m, $-42'$m$'$, 3, 32, 32$'$, 3m, 3m$'$, 6, 6mm &
\tick & \tick & \tick & \tick\\
\hline\hline
\end{tabular}
\caption{Magnetic point groups classified by the existence or absence
  of the four symmetry types of quadratic conductivities in a
  vanishing external magnetic field.}\label{tab:symm}
\end{table}

Taking the Jahn symbols in Table~\ref{tab:4-parts} as input, we have
used the {\tt MTENSOR} program~\cite{gallego-aca19} hosted on the the
Bilbao Crystallographic Server
(\url{http://www.cryst.ehu.es/cryst/mtensor}) to obtain the
symmetry-adapted forms of the four contributions to the quadratic
conductivity in each magnetic point group. The results are summarized
in Table~\ref{tab:symm}, where we indicate the existence or absence of
each contribution in each point group.  The rows of the table are
organized into four blocks: in the first block the quadratic response
is entirely absent, in the second (third) it is purely Hall-like
(Ohmic), and in the fourth both Hall and Ohmic responses are present.
Since we have not invoked specific microscopic mechanisms in setting
up Table~\ref{tab:symm}, our symmetry analysis is purely
phenomenological.  (If the last column is removed and the table is
rearranged accordingly, it reduces to the table given in
Ref.~\cite{wang-prl21}, which pertains to the three terms in
\eq{sigma-abc-tot} for the disorder-free $\sigma_\abc$.)
Interestingly, all $2^4=16$ possibilities are realized in
Table~\ref{tab:symm}. In particular, there are magnetic point groups
for which only one of the four contributions is present; clearly,
materials belonging to those point groups should be ideally suited for
studying one specific type of quadratic current response.  As already
mentioned, in the point groups where that response is purely Ohmic and
$\T$-even the quadratic current is purely disorder mediated.


\section{Discussion}
\secl{conclusions}
In this work we have shown how, given a dc conductivity tensor
of arbitrary order $n$ in the electric field, the current may be
uniquely separated into Hall and Ohmic parts
\beq
\jj^{(n)}=\jj^{(n)}_\HH+\jj^{(n)}_\OO
\eeq
by taking linear combinations of that tensor with permuted indices.
This separation is insensitive to the particular gauge choice for the
conductivity, and applying it multiple times gives the same result as
applying it only once. No other generic order-by-order partition of
the induced current fulfills these two requirements. Thus, once we
have separated the Hall and Ohmic parts we cannot make any further
subdivisions of the current into physically meaningful parts without
invoking either the symmetries of the system, or the
  microscopic processes producing the nonlinear currents.

The nonlinear Hall effect has sometimes been associated with the
transverse part of the current~\cite{lai-natnano21}, and spontaneous
unidirectional magnetoresistence with a longitudinal
response~\cite{avci-natphys15,olejnik-prb15}.  The present work
provides sharper definitions of Hall and Ohmic nonlinear responses
that are generally valid irrespective of crystal symmetry. For
example, spontaneous unidirectional magnetoresistence should be
defined as the $\T$-odd part of the quadratic Ohmic response,
which generally has both longitudinal and transverse components.  This
is consistent with the analysis in Ref.~\cite{zelezny-prb21}, where
the same conclusion was reached on the basis of a particular
mechanism, namely the nonlinear Drude term in \eq{sigma-abc-tot}.  We
hope that the present work will be useful for identifying the Hall and
Ohmic parts of nonlinear responses, both experimentally and in the
context microscopic theories.


\paragraph{Acknowledgements}
We thank David Vanderbilt for comments on the manuscript, and
Cheol-Hwan Park and Jos\'e Lu\'is Martins for discussions.

\paragraph{Funding information}
Work by S.S.T. was supported by the European Research Council (ERC)
under the European Union's Horizon 2020 research and innovation
program (ERC-StG-Neupert757867-PARATOP), and by Grant
No. PP00P2-176877 from the Swiss National Science Foundation.  Work by
I.S. was supported by Grant No.~FIS2016-77188-P from the Spanish
Ministerio de Econom\'ia y Competitividad.

\begin{appendix}

\section{Uniqueness of the partition at arbitrary order}
\secl{general-proof}
In this Appendix we prove that the Hall vs Ohmic partition of the
current described in \sref{higher-order} is the only valid generic
partition at arbitrary order $n$ in the electric field.  We start with
the general expression in \eq{Psigma} for the action of the operator
$\hat{P}$ on the conductivity,
\beq
\hat{P}\sigma_{\a_0\a_1\ldots\a_n}=
\sum_{p}\,a_p \sigma_{\a_{p(0)}\a_{p(1)}\ldots\a_{p(n)}}\,,
\eql{Psigma-2}
\eeq
where the sum is over all permutations $ \{ p(1),\ldots,p(n)\} $ of
$\{0,1,\ldots,n\}$.  The generalization of \eq{Pj-2nd} for the action
of $\hat{P}$ on the current reads
\begin{align}
\hat{P}j^{(n)}_{\a_0} &=
 \bigl(A_0\sigma_{\a_0\a_1\ldots\a_n}+A_1\sigma_{\a_1\a_0\ldots\a_n}+\ldots\nn
 &\,\,\,\,
 +A_i\sigma_{\a_1\ldots{\a_i\a_0\a_{i+1}}\ldots\a_n}+\ldots
 +A_n\sigma_{\a_1\ldots\a_n\a_0}\bigr)
E_{\a_1}\ldots E_{\a_n}\,,
\eql{Pj-n}
\end{align}
where
\beq
A_i=\longsum[10]_p^{p(i)=0}\,a_p\,.
\eql{Ai}
\eeq
Since they fully determine the projected current, the $A_i$ are the
only physically meaningful parameters, and changes in the parameters
$a_p$ that leave every $A_i$ invariant amount to gauge
transformations.

Recall from \sref{general} that a general gauge transformation
$\Delta\sigma_{\a_0\ldots\a_n}$ of the conductivity tensor must
satisfy the condition
\beq
\sum_q\,\Delta\sigma_{\a_0\a_{q(1)}\ldots\a_{q(n)}}=0\, ,
\eql{gauge-inv2}
\eeq
where the summation is over all permutations $ \{ q(1),\ldots,q(n)\} $
of $\{1,\ldots,n\}$.  As stated in \eq{gauge-inv}, we want the
projected current to be invariant under all possible gauge
transformations.  To make progress, it is sufficient to require at
this point invariance under the subset of gauge transformations
$\Delta\sigma^i_{\alpha_0\alpha_1\ldots\alpha_n}$ that are
antisymmetric under permutation of the indices at positions $i$ and
$i+1$,
\beq
\Delta\sigma^{i}_{\a_0\ldots\a_{i-1}\a_i\a_{i+1}a_{i+2}\ldots\a_n}=
-\Delta\sigma^{i}_{\a_0\ldots\a_{i-1}\a_{i+1}\a_i\a_{i+2}\ldots\a_n}\,,
\eql{Dsigma-ij}
\eeq
where $0<i<n$.  For such transformations, the gauge invariance
condition on the projected current \eqref{eq:Pj-n} takes the form
\beq
\bigl(A_i-  A_{i+1} \bigr)
\Delta\sigma^{i}_{\a_1\ldots\a_i\a_0\a_{i+1}\a_{i+2}\ldots\a_n}
E_{\a_1}\ldots E_{\a_n} =0\,.
\eeq
This condition can hold in general if and only if $A_i=A_{i+1}$, and by
letting the index $i$ run from $1$ to $n-1$ we get
\beq
A_1=A_2=\ldots=A_n\,.
\eql{gauge-inv-pairs}
\eeq
Therefore, the two parameters $A_0$ and $A_1$ fully determine the
projected current.

Let us turn now to the idempotency condition~\eqref{eq:idemp}.  Acting
with $\hat{P}$ on both sides of \eq{Psigma-2} we obtain the following
generalization of \eq{P2-2nd},
\beq
\hat{P}^2\sigma_{\a_0\a_1\ldots\a_n} =\sum_{p}\,\tilde{a}_p
\sigma_{\a_{p(0)}\a_{p(1)}\ldots\a_{p(n)}}
=\longsum[12]_{p,\,p_1,\,p_2}^{p_2\cdot p_1=p}\,a_{p_1}a_{p_2}
\sigma_{\a_{p(0)}\a_{p(1)}\ldots\a_{p(n)}}\,, \eql{Psigma-square}
\eeq
and hence the idempotency condition becomes $A_i=\tilde A_i$ for
$i=0,\ldots,n$ where, by analogy with \eq{Ai},
\bea
\tilde A_i \equiv \longsum[10]_p^{p(i)=0}\,\tilde a_i = \longsum[10]_p^{p(i)=0}
\longsum[15]_{p_1,\,p_2}^{p_2\cdot p_1=p}\,a_{p_1}a_{p_2}\,.
\eql{idemp-eq-n-1}
\eea
Solving this equation for arbitrary $n$ is not as easy as solving it
for $n=2$ [\eq{A0A1A2tilde}].  But having settled the gauge invariance
conditions in \eq{gauge-inv-pairs}, we can now pick a convenient gauge
for the coefficients $a_p$.  (This entails no loss of generality,
because we study the action of $\hat{P}$ on the physical current, not
on a particular form of the conductivity tensor.)  We choose the most
symmetric gauge compatible with \eq{gauge-inv-pairs}, namely, the
gauge where all terms in the summand of \eq{Ai} are identical,
\beq a_p=\begin{cases}
A_0/n!\,,\quad \textrm{if}\;\; p(0)=0\\
A_1/n!\,,\quad \textrm{if}\;\; p(0)\neq 0\\
\end{cases}\,.
\eeq
Substituting in \eq{idemp-eq-n-1}, the idempotency condition
$A_i=\tilde A_i$ becomes
\beq
A_i = \frac{1}{n!^2} \longsum[10]_p^{p(i)=0} 
    \left(
        \longsum[15]_{p_1,p_2}^{\tiny \begin{array}{c}p_2\cdot p_1=p\\p_1(0)=0\\ p_2(0)=0  \end{array}} A_0^2 + 
        \longsum[15]_{p_1,p_2}^{\tiny \begin{array}{c}p_2\cdot p_1=p\\p_1(0)=0\\ p_2(0)\neq 0  \end{array}} A_0 A_1 + 
        \longsum[15]_{p_1,p_2}^{\tiny \begin{array}{c}p_2\cdot p_1=p\\p_1(0)\neq 0\\ p_2(0)= 0  \end{array}} A_1 A_0 + 
        \longsum[15]_{p_1,p_2}^{\tiny \begin{array}{c}p_2\cdot p_1=p\\p_1(0)\neq 0\\ p_2(0)\neq 0  \end{array}} A_1 A_1 
    \right)\,,
\eql{idemp-eq-n-2}
\eeq
with $i$ running from $0$ to $n$. {Due to \eq{gauge-inv-pairs},}
the $n$ equations with $1\le i\le n$ are identical, leaving two
equations only. These can be written as
\beq
A_0=\frac{1}{n!^2}\left(a A_0^2+b A_0 A_1 + c A_1^2\right)\,,\quad\quad
A_1=\frac{1}{n!^2}\left(d A_0^2+e A_0 A_1 + f A_1^2\right)\,,
\eql{idemp-equations-n-gen}
\eeq
where the coefficients {$a$ to $f$} are the numbers of pairs of
permutations $p_1,p_2$ of the set $\{0,1,\ldots n\}$ that satisfy the
conditions
\begin{subequations}
\bea
a,d&:& p_1(0)=p_2(0)=0\,,\\
b,e&:& (p_1(0)=0 \;\land\; p_2(0)\neq 0)\; \lor\; 
(p_1(0)\neq 0 \;\land\; p_2(0)= 0)\,, \\
c,f&:& p_1(0)\neq 0 \;\land\; p_2(0)\neq 0\,,  
\eea
together with 
\bea
a,b,c&:& p_2(p_1(0))=0\,,\\
d,e,f&:& p_2(p_1(1))=0\,.
\eea
\end{subequations}
It now becomes a straightforward combinatorial exercise to obtain
\beq
\begin{array}{rclcrcl}
a&=&n!^2, &&d&=&0 \\
b&=&0,&&e&=& 2\cdot n!^2\\
c&=&n \cdot n!^2, && f&=&  (n-1)\cdot n!^2
\end{array}\,,
\eeq
which leads to the following generalization of \eq{idemp-equations-2nd},
\beq
A_0=A_0^2+n A_1^2\,,\quad\quad
A_1=2A_0 A_1+(n-1)A_1^2\,.
\eeq
Apart from the trivial solutions $\hat{P}_0$ and $\hat{P}_1$ of the
same type as in \eq{trivial}, these equations have the solutions
\beq
\begin{cases}
\hat{P}_\HH:  (A_0,A_1=\ldots=A_n)=\left(\frac{n}{n+1},-\frac{1}{n+1}\right)\\
\hat{P}_\OO:  (A_0,A_1=\ldots=A_n)=\left(\frac{1}{n+1},\frac{1}{n+1}\right)
\end{cases}\,,
\eql{nontrivial-N}
\eeq
which generalize \eq{nontrivial}. It can be readily verified that the
solution for $\hat{P}_\OO$ is satisfied by \eq{POn}. And since
$\hat{P}_\HH$ fulfills the Hall condition~\eqref{eq:Hall-like-2nd} but
$\hat{P}_\OO$ does not, we have obtained a unique partition of the
$n$-th order current into Hall and Ohmic components.


\section{Repackaging of the disorder-free quadratic Hall conductivity}
\secl{repackaging}
Inserting \eq{sigma-abc-tot} for $\sigma_\abc$ in \eq{chi-H} for
$\chi^\HH_{\c\d}$ and writing
$\Omega_n^{\a\b}=\varepsilon_{\a\b\c}\Omega_n^\c$ one finds
\beq
\chi^\HH_{\a\b}=\frac{e^3}{\hbar}\int_{\kk n}f_0(\epsilon_n)
\varepsilon_{\a\c\d}\partial_\c G_n^{\d\b}+
\frac{e^3\tau}{\hbar^2}
\left[ D_{\b\a}-\frac{1}{3}\delta_{\a\b}\text{Tr}(D)\right]\,,
\eql{chi-H-berry-a}
\eeq
where
\beq
D_{\b\a}=\int_{\kk n}f_0(\epsilon_n)\partial_\b\Omega_n^\a
\eeq
is the Berry curvature dipole~\cite{sodemann-prl15}.  The first in
\eq{chi-H-berry-a} agrees with the expression obtained in
Ref.~\cite{liu-prl21} starting from the gauge-dependent definition
$\chi^\HH_{\c\d}=\varepsilon_{\a\b\c}\sigma_{\a\b\d}/2$. The second
term agrees with the expression in Eq.~(8) of Ref.~\cite{nandy-prb19},
once that expression is multiplied by the factor of 4/3 that was
discussed in connection with
\eqs{sigma-1-2-tilde}{sigma-1-3-tilde}. That second term can be
simplified by noting that $\text{Tr}(D)=0$ for topological
reasons~\cite{tsirkin-prb18,konig-prb19}, yielding
\beq
\chi^\HH_{\a\b}=\frac{e^3}{\hbar}\int_{\kk n}f_0(\epsilon_n)
\left[
\varepsilon_{\a\c\d}\partial_\c G_n^{\d\b}+
(\tau/\hbar)\partial_\b\Omega_n^\a
\right]
\eql{chi-H-berry}
\eeq
for the disorder-free quadratic Hall tensor. The first term is $\T$
odd and intrinsic (independent of $\tau$), and the second is $\T$ even
and extrinsic.



\end{appendix}




\bibliography{bib}

\begin{thebibliography}{10}
\providecommand{\url}[1]{\texttt{#1}}
\providecommand{\urlprefix}{URL }
\expandafter\ifx\csname urlstyle\endcsname\relax
  \providecommand{\doi}[1]{doi:\discretionary{}{}{}#1}\else
  \providecommand{\doi}{doi:\discretionary{}{}{}\begingroup
  \urlstyle{rm}\Url}\fi
\providecommand{\eprint}[2][]{\url{#2}}

\bibitem{baranova-oc77}
N.~B. Baranova, Y.~V. Boddanov and B.~Y. Zel'Dovich,
\newblock \emph{{Electrical analog of the Faraday effect and other new optical
  effects in liquids}},
\newblock Optics Commun. \textbf{22}, 243 (1977),
\newblock \doi{10.1016/0030-4018(77)90028-1}.

\bibitem{ivchenko-jetp78}
E.~L. Ivchenko and G.~E. Pikus,
\newblock \emph{{New photogalvanic effect in gyrotropic crystals}},
\newblock JETP Lett. \textbf{27}, 604 (1978).

\bibitem{vorobev-jetp79}
E.~L. Vorob'ev, E.~L. Ivchenko, G.~E. Pikus, I.~I. Farbshtein, V.~A. Shalygin
  and A.~V. Shturbin,
\newblock \emph{{Optical activity in tellurium induced by a current}},
\newblock JETP Lett. \textbf{29}, 441 (1979).

\bibitem{tokura-nc18}
Y.~Tokura and N.~Nagaosa,
\newblock \emph{{Nonreciprocal responses from non-centrosymmetric quantum
  materials}},
\newblock Nat. Commun. \textbf{9}, 3740 (2018),
\newblock \doi{10.1038/s41467-018-05759-4}.

\bibitem{rikken-prl01}
G.~L. J.~A. Rikken, J.~F\"olling and P.~Wyder,
\newblock \emph{Electrical magnetochiral anisotropy},
\newblock Phys. Rev. Lett. \textbf{87}, 236602 (2001),
\newblock \doi{10.1103/PhysRevLett.87.236602}.

\bibitem{ideue-natphys17}
T.~Ideue, K.~Hamamoto, S.~Koshikawa, M.~Ezawa, S.~Shimizu, Y.~Kaneko,
  Y.~Tokura, N.~Nagaosa and Y.~Iwasa,
\newblock \emph{{Bulk rectification effect in a polar semiconductor}},
\newblock Nat. Phys. \textbf{13}, 578 (2017),
\newblock \doi{10.1038/nphys4056}.

\bibitem{Rikken-Te}
G.~L. J.~A. Rikken and N.~Avarvari,
\newblock \emph{Strong electrical magnetochiral anisotropy in tellurium},
\newblock Phys. Rev. B \textbf{99}, 245153 (2019),
\newblock \doi{10.1103/PhysRevB.99.245153}.

\bibitem{avci-natphys15}
C.~O. Avci, K.~Garello, A.~Ghosh, M.~Gabureac, F.~A. Santos and P.~Gambardella,
\newblock \emph{{Unidirectional spin Hall magnetoresistance in
  ferromagnet/normal metal bilayers}},
\newblock Nat. Phys. \textbf{11}, 570 (2015),
\newblock \doi{10.1038/nphys3356}.

\bibitem{olejnik-prb15}
K.~Olejn\'{\i}k, V.~Nov\'ak, J.~Wunderlich and T.~Jungwirth,
\newblock \emph{Electrical detection of magnetization reversal without
  auxiliary magnets},
\newblock Phys. Rev. B \textbf{91}, 180402 (2015),
\newblock \doi{10.1103/PhysRevB.91.180402}.

\bibitem{zelezny-prb21}
J.~Zelezn\'y, Z.~Fang, K.~Olejn\'{\i}k, J.~Patchett, F.~Gerhard, C.~Gould,
  L.~W. Molenkamp, C.~Gomez-Olivella, J.~Zemen, T.~Tich\'y, T.~Jungwirth and
  C.~Ciccarelli,
\newblock \emph{{Unidirectional magnetoresistance and spin-orbit torque in
  NiMnSb}},
\newblock Phys. Rev. B \textbf{104}, 054429 (2021),
\newblock \doi{10.1103/PhysRevB.104.054429}.

\bibitem{deyo2009semiclassical}
E.~Deyo, L.~E. Golub, E.~L. Ivchenko and B.~Spivak,
\newblock \emph{Semiclassical theory of the photogalvanic effect in
  non-centrosymmetric systems},
\newblock \doi{10.48550/ARXIV.0904.1917} (2009), \eprint{0904.1917}.

\bibitem{moore-prl10}
J.~E. Moore and J.~Orenstein,
\newblock \emph{{Confinement-Induced Berry Phase and Helicity-Dependent
  Photocurrents}},
\newblock Phys. Rev. Lett. \textbf{105}, 026805 (2010),
\newblock \doi{10.1103/PhysRevLett.105.026805}.

\bibitem{gao-prl14}
Y.~Gao, S.~A. Yang and Q.~Niu,
\newblock \emph{{Field Induced Positional Shift of Bloch Electrons and Its
  Dynamical Implications}},
\newblock Phys. Rev. Lett. \textbf{112}, 166601 (2014),
\newblock \doi{10.1103/PhysRevLett.112.166601}.

\bibitem{sodemann-prl15}
I.~Sodemann and L.~Fu,
\newblock \emph{{Quantum Nonlinear Hall Effect Induced by Berry Curvature
  Dipole in Time-Reversal Invariant Materials}},
\newblock Phys. Rev. Lett. \textbf{115}, 216806 (2015),
\newblock \doi{10.1103/PhysRevLett.115.216806}.

\bibitem{nandy-prb19}
S.~Nandy and I.~Sodemann,
\newblock \emph{{Symmetry and quantum kinetics of the nonlinear Hall effect}},
\newblock Phys. Rev. B \textbf{100}, 195117 (2019),
\newblock \doi{10.1103/PhysRevB.100.195117}.

\bibitem{ma-nature19}
Q.~Ma, S.-Y. Xu, H.~Shen, D.~MacNeill, V.~Fatemi, T.-R. Chang, A.~M.~M.
  Valdivia, S.~Wu, Z.~Du, C.-H. Hsu, S.~Fang, Q.~D. Gibson \emph{et~al.},
\newblock \emph{{Observation of the nonlinear Hall effect under
  time-reversal-symmetric conditions}},
\newblock Nature \textbf{565}, 337 (2019),
\newblock \doi{10.1038/s41586-018-0807-6}.

\bibitem{kang-natmater19}
K.~Kang, T.~Li, E.~Sohn, J.~Shan and K.~F. Mak,
\newblock \emph{{Nonlinear anomalous Hall effect in few-layer WTe$_2$}},
\newblock Nat. Mater. \textbf{18}, 324 (2019),
\newblock \doi{10.1038/s41563-019-0294-7}.

\bibitem{li-prb21}
R.-H. Li, O.~G. Heinonen, A.~A. Burkov and S.~S.-L. Zhang,
\newblock \emph{{Nonlinear Hall effect in Weyl semimetals induced by chiral
  anomaly}},
\newblock Phys. Rev. B \textbf{103}, 045105 (2021),
\newblock \doi{10.1103/PhysRevB.103.045105}.

\bibitem{ortix2021nonlinear}
C.~Ortix,
\newblock \emph{{Nonlinear Hall Effect with Time-Reversal Symmetry: Theory and
  Material Realizations}},
\newblock Adv. Quantum Technol. \textbf{4}, 2100056 (2021),
\newblock \doi{10.1002/qute.202100056}.

\bibitem{du2021perspective}
Z.~Z. Du, H.-Z. Lu and X.~C. Xie,
\newblock \emph{{Nonlinear Hall Effects}},
\newblock Nat. Rev. Phys. \textbf{3}, 744 (2021),
\newblock \doi{10.1038/s42254-021-00359-6}.

\bibitem{zhang2021higherorder}
C.-P. Zhang, X.-J. Gao, Y.-M. Xie, H.~C. Po and K.~T. Law,
\newblock \emph{{Higher-Order Nonlinear Anomalous Hall Effects Induced by Berry
  Curvature Multipoles}},
\newblock \doi{10.48550/ARXIV.2012.15628} (2021), \eprint{2012.15628}.

\bibitem{wang-prl21}
C.~Wang, Y.~Gao and D.~Xiao,
\newblock \emph{{Intrinsic Nonlinear Hall Effect in Antiferromagnetic
  Tetragonal CuMnAs}},
\newblock Phys. Rev. Lett. \textbf{127}, 277201 (2021),
\newblock \doi{10.1103/PhysRevLett.127.277201}.

\bibitem{liu-prl21}
H.~Liu, J.~Zhao, Y.-X. Huang, W.~Wu, X.-L. Sheng, C.~Xiao and S.~A. Yang,
\newblock \emph{{Intrinsic Second-Order Anomalous Hall Effect and Its
  Application in Compensated Antiferromagnets}},
\newblock Phys. Rev. Lett. \textbf{127}, 277202 (2021),
\newblock \doi{10.1103/PhysRevLett.127.277202}.

\bibitem{lai-natnano21}
S.~Lai, H.~Liu, Z.~Zhang, J.~Zhao, X.~Feng, N.~Wang1, C.~Tang1, Y.~Liu, K.~S.
  Novoselov, S.~A. Yang and W.~bo~Gao,
\newblock \emph{{Third-order nonlinear Hall effect induced by the
  Berry-connection polarizability tensor }},
\newblock Nat. Nanotechnol. \textbf{16}, 869 (2021),
\newblock \doi{10.1038/s41565-021-00917-0}.

\bibitem{boyd-book03}
R.~W. Boyd,
\newblock \emph{Nonlinear Optics},
\newblock Academic Press, 2nd edn. (2003).

\bibitem{gao-fp19}
Y.~Gao,
\newblock \emph{{Semiclassical dynamics and nonlinear charge current}},
\newblock Front. Phys. \textbf{14}, 33404 (2019),
\newblock \doi{10.1007/s11467-019-0887-2}.

\bibitem{SHTRIKMAN1965147}
S.~Shtrikman and H.~Thomas,
\newblock \emph{Remarks on linear magneto-resistance and
  magneto-heat-conductivity},
\newblock Solid State Commun. \textbf{3}, 147 (1965),
\newblock \doi{10.1016/0038-1098(65)90178-X},
\newblock Erratum: \textbf{3} civ. (1965).

\bibitem{grimmer1993}
H.~Grimmer,
\newblock \emph{{General relations for transport properties in magnetically
  ordered crystals}},
\newblock Acta Crystallogr. A \textbf{49}, 763 (1993),
\newblock \doi{10.1107/S0108767393003770}.

\bibitem{xiao-rmp10}
D.~Xiao, M.-C. Chang and Q.~Niu,
\newblock \emph{Berry phase effects on electronic properties},
\newblock Rev. Mod. Phys. \textbf{82}, 1959 (2010),
\newblock \doi{10.1103/RevModPhys.82.1959}.

\bibitem{isobe-sciadv20}
H.~Isobe, S.-Y. Xu and L.~Fu.,
\newblock \emph{{High-frequency rectification via chiral Bloch electrons}},
\newblock Sci. Adv. \textbf{6}, eaay2497 (2020),
\newblock \doi{10.1126/sciadv.aay2497}.

\bibitem{konig-prb19}
E.~J. K\"onig, M.~Dzero, A.~Levchenko and D.~A. Pesin,
\newblock \emph{{Gyrotropic Hall effect in Berry-curved materials}},
\newblock Phys. Rev. B \textbf{99}, 155404 (2019),
\newblock \doi{10.1103/PhysRevB.99.155404}.

\bibitem{du-natcomms19}
Z.~Z. Du, C.~M. Wang, S.~Li, H.-Z. Lu and X.~Xie,
\newblock \emph{{Disorder-induced nonlinear Hall effect with time-reversal
  symmetry}},
\newblock Nat. Commun. \textbf{10}, 3047 (2019),
\newblock \doi{10.1038/s41467-019-10941-3}.

\bibitem{jahn-actacrys49}
H.~A. Jahn,
\newblock \emph{{Note on the Bhagavantam-Suranarayana method of enumerating the
  physical constants of crystals}},
\newblock Acta Cryst. \textbf{2}, 30 (1949),
\newblock \doi{10.1107/S0365110X49000060}.

\bibitem{gallego-aca19}
S.~V. Gallego, J.~Etxebarria, L.~Elcoro, E.~S. Tasci and J.~M. Perez-Mato,
\newblock \emph{{Automatic calculation of symmetry-adapted tensors in magnetic
  and non-magnetic materials: a new tool of the Bilbao Crystallographic
  Server}},
\newblock Acta Cryst. A \textbf{75}, 438 (2019),
\newblock \doi{10.1107/S2053273319001748}.

\bibitem{tsirkin-prb18}
S.~S. Tsirkin, P.~A. Puente and I.~Souza,
\newblock \emph{Gyrotropic effects in trigonal tellurium studied from first
  principles},
\newblock Phys. Rev. B \textbf{97}, 035158 (2018),
\newblock \doi{10.1103/PhysRevB.97.035158}.

\end{thebibliography}

\nolinenumbers

\end{document}